\documentclass[11pt,letterpaper]{article}
\usepackage[margin=1in]{geometry}
\usepackage[T1]{fontenc}
\usepackage{lmodern}
\usepackage{amsmath,amssymb,amsthm}
\usepackage{booktabs,array,graphicx}
\usepackage[numbers,sort&compress]{natbib}
\usepackage[hidelinks]{hyperref}
\newtheorem{proposition}{Proposition}
\newcommand{\NAtomicInstances}{26}
\newcommand{\NAtomicPairs}{676}
\newcommand{\NAtomicProofs}{1562}
\newcommand{\NAtomicQueries}{2028}
\newcommand{\NAtomicWitnesses}{466}
\newcommand{\NBrowserCases}{3}
\newcommand{\NCheckerDecisions}{312}
\newcommand{\NCheckerHistories}{156}
\newcommand{\NChromaCases}{18}
\newcommand{\NExecutedComponents}{5}
\newcommand{\NFilesystemCases}{18}
\newcommand{\NFrameComponents}{6}
\newcommand{\NFrameRepositories}{4}
\newcommand{\NGitCases}{45}
\newcommand{\NIncompleteReplyCases}{27}
\newcommand{\NIntentRejections}{20}
\newcommand{\NMemoryCases}{9}
\newcommand{\NMicroPairs}{12}
\newcommand{\NMicroProofs}{15}
\newcommand{\NMicroQueries}{24}
\newcommand{\NMicroWitnesses}{9}
\newcommand{\NNativeCases}{60}
\newcommand{\NPhysicalRejections}{6}
\newcommand{\NPrefixBlindFailures}{3}
\newcommand{\NPrefixCases}{6}
\newcommand{\NPrimaryExecutions}{186}
\newcommand{\NUnrealizedPrimarySchedules}{2}

\title{Checking Retry Transparency in Agent Tools}
\author{Jiapeng Li\\Microsoft}
\date{November 11, 2026}
\begin{document}
\maketitle

\begin{abstract}
A successful retry does not necessarily put the earlier attempt behind it.
An old write may still overwrite a later acknowledged update; a partially
executed file write may even empty a file after the retry has returned
successfully. We study how to test the recovery guarantee behind an agent
tool's local idempotence claim. Our prototype takes manually constructed,
source-backed effect contracts and checks them at two levels: atomic
retry-absorption obligations and bounded interleavings of successful-attempt
microsteps. It emits model witnesses, conditional admission decisions, and
explicit uncertainty when a completion boundary cannot be justified.
The evaluation preserves a locally frozen purposive frame of
\NFrameComponents{} MCP components from \NFrameRepositories{} repositories;
selected paths in \NExecutedComponents{} execute on the study machine and one
backend remains unavailable. Real tests include file, graph, Git-index,
document-update and browser bindings, together with a native Qdrant development
matrix. In \NIncompleteReplyCases{} missing-reply checks, the selective rule
completes six justified workflows without an acknowledged goal violation,
whereas strict coalescing completes none and blind retry completes nine with
three violations. An intent-aware Porcupine specification independently agrees
with all \NCheckerHistories{} complete-history oracles. These are
mechanism-selected tests, not defect-prevalence estimates or a new
linearizability algorithm. The contribution is an executable recovery-contract
workflow that makes effect granularity, logical identity and completion
assumptions testable rather than inheriting them from a tool annotation.
\end{abstract}

\section{Introduction}
The file was correct when the retry returned. It was also correct when every
physical call had finished. Between those observations, however, it was empty.
The original write had opened the file before the deadline, paused before
truncation, and resumed after the successful retry. Its truncate erased the
acknowledged content; its later write restored it. Both calls eventually
returned normally. A test that inspected only the settled file would miss
the intervening loss of an acknowledged state.

There is a simpler version of the problem even when each write is atomic.
Suppose logical operation \(A\) writes 10 to an object. A fast retry completes
\(A\), after which logical operation \(B\) writes 20. If an older physical
attempt of \(A\) subsequently writes 10, the object count remains one but the
acknowledged \(A;B\) workflow is no longer represented by its state. Reusing an
object identifier has not created an operation-deduplication protocol.

Neither observation overturns distributed-systems theory. Duplicate
suppression is a longstanding RPC concern \citep{birrell1984}; RIFL explains
why individually idempotent writes are not interchangeable with exactly-once
linearizable execution \citep{rifl}. Linearizability already provides the
appropriate machinery for reasoning about many concurrent-object histories
\citep{herlihy1990}. The practical question here is what a recovery test may
conclude from the contracts available at an agent-tool boundary.

MCP tools expose invocation schemas, results and annotations such as
\texttt{idempotentHint}. These are useful descriptions, not implementations
of a business-operation ledger \citep{mcpschema}. A tool's local repeatability
does not identify the logical operation being repeated, specify the granularity
of its effects, or establish that a normal RPC reply exhausts an application's
future work. Existing fault-injection and conformance tools already examine
substantial portions of this space \citep{mcprig,mcpinjector,mcpguard,toolcontracts}.
Another collection of dropped replies would therefore be insufficient as a
method contribution.

We instead develop a recovery-contract workflow. Given a declared logical
workflow and a source-backed effect model, the prototype checks whether an
outstanding older attempt can invalidate an acknowledged state, constructs
candidate witnesses, and selects a conditional action: admit, await a physical
attempt, coalesce with it, or report ambiguity. The method does not infer a
business intent or an entire implementation model from an MCP schema.
Source-to-model bindings remain manual and reviewable.

The paper contributes three things. First, it separates logical-operation
identity, effect granularity and observable completion in one executable
contract. Second, it connects atomic algebraic checks and bounded microstep
checks to actual I/O-delay adapters and an explicit admission policy. Third,
it evaluates the resulting claims with retained negative and unavailable
targets, independently coded interpreters and an actually executed
linearizability checker. The atomic laws, concurrency enumeration and
UNKNOWN outcome are not individually claimed as new.

\section{What is being guaranteed?}
\label{sec:scope}
\subsection{Logical operations and physical attempts}
An application supplies a sequence of distinct logical intents
\(L=(\ell_1,\ldots,\ell_n)\). An intent can have several physical attempts.
Two attempts with the same arguments are not automatically the same intent;
two independently requested operations remain distinct even when their
payloads match. In our experiments intent labels are supplied by the
controlled workflow, not inferred from language or argument equality.

For a declared initial state \(s\), the fault-free reference for two intents
is \(F_B(F_A(s))\), written \(A;B(s)\) with execution proceeding left to right.
The reference comes from the intended selected API operations. For example,
two first-wins inserts into the same existing-name domain need not produce
two entities, whereas a reset followed by an old stage operation must not
silently be treated as the acknowledged stage--reset workflow.

The actual history includes the invocation and return intervals of physical
calls, their association with intents, and independently observed state.
The selected state signature is explicit: exact file bytes, graph entities,
Git-index blobs, or documents read by ID. Approximate retrieval quality,
natural-language planning quality and arbitrary external clients are outside
the study.

\subsection{Settled state is weaker than an acknowledged prefix}
We use two observations, not one interchangeable success flag.
\emph{Settled effect agreement} asks whether state after the tracked attempts
stop agrees with the canonical logical workflow. \emph{Acknowledged-prefix
stability} additionally asks whether an older attempt can invalidate that
state after the foreground operation has acknowledged completion.

The prefix obligation is checked after the foreground acknowledgement and
after each remaining older effect step. It does not merely compare two final
record counts. Nor does it assert complete API linearizability for every
unobserved read, nondeterministic return value or concurrent external caller.
When we use a whole-tool atomic specification in an independent checker, a
rejection denotes failure of that specified refinement, not proof that the
tool promised atomicity.

\subsection{Acknowledgement, quiescence and progress}
A usable acknowledgement is projected to a normal unit result for the selected
effect operation. We do not certify replay of the original receipt, operation
identifier, response body or billing record. These projections matter:
an old Git staging call can report a different message from a fresh no-op
staging call even if their index effects agree.

Strong completion evidence must establish the relevant effect boundary.
For an awaited descriptor write and close, a normal tool reply follows those
steps. For a browser click that schedules a future application timer, the
normal click reply establishes neither the application's durable write nor
quiescence of that future work. A timeout, cancellation request or failed
connection is not promoted to strong completion.

Finally, blocking is not successful recovery. A policy that never acknowledges
\(B\) may avoid a violated \(B\) acknowledgement while failing to complete the
workflow. We report logical completion and effect agreement separately,
especially when the original reply is permanently unavailable.

\section{Recovery contracts and the prototype}
\subsection{An explicit, limited representation}
A profile records the selected operations, resource mapping, state
observations, invariant, intent labels, effect granularity, completion binding,
and source/version evidence. The atomic IR contains immutable typed state
fields and simultaneous operation updates. Its library includes overwrite,
first-wins insert, deletion, monotone maximum and counter-update instances.
It is a contract language, not an automatic source-code extractor.

An observation projection is safe for compositional rewrites only if future
declared operations cannot reveal a discarded difference. The implementation
therefore retains the backward dependency closure of observations and the
invariant over every operation in the declared context. A regression with a
hidden phase field and a later reveal operation demonstrates why equality of
the immediately visible field is insufficient.

The initial invariant must be satisfiable and preserved by the declared
operations before compositional laws are used. An empty domain cannot create
a vacuous certificate. Operations outside the context are rejected. Chroma's
update-only profile, for instance, explicitly requires both selected document
IDs to exist; the real setup and exact-get checks establish that test domain.
It would be incorrect to model an ignored update to an absent ID as creation.

\subsection{Atomic obligations}
For instantiated operations \(A\) and \(B\), the checker asks whether a
counterexample state exists for:
\begin{align}
 A;A &\ne A, &&\text{isolated idempotence},\\
 A;B &\ne B;A, &&\text{commutativity},\\
 A;B;A &\ne A;B, &&\text{ordered retry absorption}.
\label{eq:absorption}
\end{align}
Queries are evaluated over the retained state and invariant. SAT yields a model
witness; UNSAT yields \texttt{PROVED\_MODEL}, conditional on the contract;
solver UNKNOWN never becomes PASS. Every exported query is hash-bound to its
actual UTF-8 bytes and can be reloaded.

Absorption can admit an operation that a commutativity rule would fence.
Let \(A\) insert 10 if a key is absent and \(B\) insert 20 under the same
condition. On an initially absent key, \(A;B\) and \(B;A\) differ. Yet a later
repeat of \(A\) cannot change \(A;B\), provided deletion, expiry and other
invalidating operations are absent from this context. Replacing \(B\) with
delete removes that argument: \(A;B;A\) can resurrect the key.

\begin{proposition}[A standard absorption consequence]
Suppose the invariant domain is closed under the operations. If
\(A;A=A\) and \(A;B_i;A=A;B_i\) for each \(B_i\) in a later operation
sequence, then \(A;B_1;\cdots;B_k;A=A;B_1;\cdots;B_k\).
\end{proposition}
\begin{proof}
For \(k=0\), use idempotence. For the inductive step, write
\(W=B_1;\cdots;B_{k-1}\). The induction hypothesis permits insertion of \(A\)
after \(A;W\):
\[
 A;W;B_k;A=A;W;A;B_k;A=A;W;A;B_k=A;W;B_k.
\]
The equalities are universal on the invariant domain, so they remain valid in
the declared closed context.
\end{proof}
This is established algebraic reasoning, closely related to left regular band
identities and mover analysis, not a new mathematical theorem. Existing
commutativity and left/right-mover inference is substantially more general
than a few new named SMT queries \citep{servois}. Our use of the obligation is
to govern tracked recovery attempts under explicit bindings.

\subsection{Why whole-tool atomicity must be checked}
The first filesystem source inspection refuted the intended atomic-overwrite
abstraction. The implementation preserves an existing file's identity by
truncating and then writing through a descriptor. These are separate awaited
steps. An older short write can resume after a newer long write and leave a
prefix from the old value followed by the new value's suffix.

The original atomic analysis reports unsupported atomicity for that profile.
It does not issue a concrete certificate and then reinterpret a failing test
as an upstream violation. This source/model mismatch motivated a separate
bounded microstep engine.

\subsection{Bounded microsteps and post-acknowledgement checks}
The microstep vocabulary includes register updates, first-wins insert and
delete, taking and publishing a modified snapshot, separate truncate and
prefix-write steps, and lock acquisition and release.
It distinguishes shared fields from attempt-local snapshots. Lock ownership
constrains which schedules are feasible; malformed or unreleased locks cannot
produce an empty-domain certificate.

For one old \(A\), one retry of \(A\), and optional successor \(B\), the
foreground chain completes the retry before invoking \(B\). The engine
enumerates all interleavings of that chain with the old attempt's ordered
steps. For each feasible schedule it asks whether any selected state after
the foreground acknowledgement, including any remaining old-step prefix,
differs from the canonical reference. For the retry-only query the reference
is \(A\); for the successor query it is \(A;B\).

UNSAT is labeled \texttt{PROVED\_BOUNDED\_MODEL}, not a theorem about an
arbitrary number of retries or all implementation executions. The implemented
byte model has an explicit resource footprint and byte-width bound. The
filesystem prediction uses a two-byte write and a 16-byte bound; it does not
justify atomicity of arbitrary filesystem writes. The instruction boundaries
remain source-model assumptions that must be materialized and checked.

This distinction changes a recovery action. A repeated truncate/write can
invalidate an acknowledged retry before any \(B\) is considered, even if its
settled result is idempotent. The microstep decision is then to await the old
call before acknowledging a retry, not merely to fence the successor.
Conversely, an unlocked snapshot/modify/publish model can lose a disjoint
later update that atomic per-key overwrites would preserve. A real critical
section removes those snapshot interleavings. These snapshot cases calibrate
the model engine; they are not reported as discovered Git defects.

\subsection{Admission and uncertainty}
The atomic runtime monitor retains physical attempts after the first logical
success. A normal successful retry does not retire another outstanding attempt.
For a new logical operation, it either proves the required absorption
obligations, awaits strongly observable conflicting attempts, or returns
ambiguity. Same-intent parameters are immutable, and a completed intent cannot
start a fresh physical retry.

The bounded decision procedure first checks retry-prefix stability. If it
fails, strong observable completion permits waiting for the old call and
coalescing with its result; without that evidence, the result is UNKNOWN. If
the retry passes, the procedure checks the successor and, where necessary, the
corresponding fenced schedules. Its scope is the declared two- or three-attempt
problem, not an unbounded workflow controller.

A failed original transport can therefore remain ambiguous while the monitor
admits a provably neutral retry and successor. No private fault-phase label,
backend acknowledgement observed only by the injector, or known delay bound
is supplied to that decision. An unsafe conflict without observable
quiescence remains ambiguous. The prototype's outputs retain
\texttt{model\_only} and explicitly deny an exact-receipt certificate.

\subsection{Witness binding and validation}
Countermodels and schedules are candidate test instructions, not implementation
defects. Adapters map selected steps to public runtime boundaries: an HTTP
request, a descriptor truncate or write, a native Git command, or the real
Chroma SDK update. They delay the original function or request without changing
its input or substituting a successful result. State is read independently of
the tool wrapper.

Each scored execution must establish its actual happens-before order.
Serialized, lock-protected event ordinals avoid relying on equal timestamps;
the Windows Python clock used in an early pilot had insufficient resolution
for strict timestamp comparisons. Unrealized schedules are retained. The
runtime guard receives contracts and allowed completion evidence, not the
injector's private placement of a pause.

\section{Study design}
\subsection{A frozen frame, with an explicit refinement boundary}
Before new mutation-source inspection, we locally froze six purposively chosen
components and exact commits. Table~\ref{tab:frame} gives the frame and selected
paths. It is not a random sample, and the timestamp/hash record is not an
external preregistration.

\begin{table}[t]
\centering\small
\begin{tabular}{p{0.20\linewidth}p{0.22\linewidth}p{0.25\linewidth}p{0.21\linewidth}}
\toprule
Component & Pinned version & Selected path & Study disposition\\
\midrule
Reference filesystem & 2.2.0 & \texttt{write\_file} & Initial frame; microstep development\\
Reference memory & 2.1.0 & Create/delete entities & Initial frame; negative control and development\\
Reference Git & 2026.11.4 & Stage/reset index & Prospective after microstep freeze\\
Chroma MCP & 0.2.6; backend 1.0.16 & Existing-document update & Prospective after microstep freeze\\
Playwright MCP & 0.0.85; core alpha 2026-11-09 & Browser click & Prospective binding; controlled application\\
Redis MCP & 0.5.1 & No recovery path executed & Backend unavailable; retained\\
\bottomrule
\end{tabular}
\caption{The original six-component frame from four repositories
\citep{reference,chroma,playwright,redis}. The three reference components share
a repository and must not be counted as three independent projects.}
\label{tab:frame}
\end{table}

Filesystem and memory were already known when the microstep refinement was
designed. They consequently become development evidence for that refinement.
The remaining original candidates---Git, Chroma, Playwright and Redis---were
not replaced. Their mutation implementations were opened after the refined
core was frozen, and each concrete profile and oracle was frozen before its
recovery executions. This separation limits, rather than eliminates, the
researcher's influence on the evaluation.

Chroma's selected commit dates from 2025. It is a historical pinned target,
not evidence of recently maintained-server behavior. Redis could not execute
locally: the installed Docker Linux daemon was unavailable and no native
Redis-compatible server was on PATH. We did not start a global service or
substitute another product to obtain a successful row.

\subsection{Concrete executions and controls}
The study contains \NPrimaryExecutions{} primary executions or binding probes,
split into deliberately different cohorts:
\NNativeCases{} native atomic development checks,
\NIncompleteReplyCases{} native missing-reply checks,
\NFilesystemCases{} filesystem recovery checks,
\NPrefixCases{} filesystem prefix checks,
\NMemoryCases{} graph checks,
\NGitCases{} Git checks,
\NChromaCases{} document checks and
\NBrowserCases{} browser-lifetime probes. This total is not an inferential
sample size: the cohorts test different predicates, and repeated
mechanism-selected cells are not independent systems.

The native Qdrant development backend is version 1.20.0. It uses unique
point namespaces, exact state reads and actual HTTP delivery boundaries.
We test overwrite conflict, first-wins conflict, resurrection, disjoint
overwrite and same-value overwrite. It is an already known development
anchor, not an extra unseen MCP component.

The MCP paths use real unmodified backing implementations. Filesystem tests
read actual temporary-file bytes. Memory tests read persisted graph JSONL.
Git tests read index blobs using separate native Git commands. Two tracked MCP
processes share one new repository, representing an original connection that
may remain active when a replacement connection opens. No commits or pushes
occur during recovery. Chroma uses two tracked MCP processes and a separate
loopback native server; direct SDK exact-get calls observe documents. Both
document IDs exist throughout, and every update supplies deterministic
three-dimensional vectors. No paid embedding API or retrieval-quality
evaluation is involved.

Fault-free functional controls precede recovery tests. Fresh supplemental
filesystem reexecutions additionally inspect the exact \(A;B\) state before
the \(A;B;A\) check, so a correct final \(A\) cannot mask an incorrect \(B\).
Those six reexecutions are supplemental validation, not extra independent
systems or part of the primary total.

Application deadlines leave the original physical call outstanding rather
than relabeling cancellation as completion. The injector releases an old call
after \(B\)'s acknowledgement or at a fixed fair-delay bound. The same
rule applies to the policies being compared. In the missing-reply cohort the
old request actually reaches the backend, but its normal response is
suppressed and the client observes a real transport failure. That private
backend completion is used for validation, never as guard input.

\subsection{Compared rules and independent checks}
The executed policy arms are idempotence-only admission, a full successor
fence, a same-contract commutativity rule, absorption-based admission, and
strict in-flight coalescing where included. These are named rules in our
prototype, not claims that we executed or outperformed Servois, RIFL or
EffectFence. In-flight coalescing is important: simply refusing a new call
until an old result arrives is often the appropriate existing solution.

For complete histories we also execute Porcupine 1.3.1 \citep{porcupine},
with two specifications. The physical-API specification treats every
physical write as its own operation. The intent-aware specification retains
distinct logical intents but suppresses another state transition for a
previously executed intent. Both receive the same unit-acknowledgement and
state observation projection. A physical pass and an intent-aware rejection
answer different questions; the former is not a defect in the checker.

The complete-history corpus contains \NCheckerHistories{} histories.
We do not fabricate a normal return for the suppressed response, so the
missing-reply cohort is reported separately rather than added to this
complete-history comparison. The browser probe likewise tests an
application-effect completion binding, not a promised atomic click-to-database
operation.

\subsection{Calibration of the model engines}
The atomic calibration contains \NAtomicInstances{} primitive instances,
\NAtomicPairs{} ordered pairs and \NAtomicQueries{} law comparisons.
All \NAtomicProofs{} model proofs agree with a separate finite concrete
cross-check; all \NAtomicWitnesses{} countermodels replay in a separately
implemented primitive interpreter. These are model checks, not service
experiments or independent human review.

The microstep calibration contains \NMicroPairs{} semantic pairs and
\NMicroQueries{} comparisons: \NMicroProofs{} bounded proofs and
\NMicroWitnesses{} independently replayed countermodels. A second concrete
interpreter enumerates schedules using a different construction and agrees
on all comparisons. Separate regressions cover hidden-state closure,
invariant feasibility and preservation, solver UNKNOWN, simultaneous updates,
late-reply tracking, query hashes, lock feasibility and Windows process
cleanup. Byte-distinct SMT files are not counted as independent laws.

\section{Results}
\subsection{Atomic idempotence does not settle interference}
Table~\ref{tab:native} shows the native development matrix.
The idempotence-only rule acknowledges all selected workflows but preserves
the wrong settled state in five executions. One additional intended conflict
schedule releases early and is retained as unrealized, not taken as evidence
that the unsafe rule is safe. The absorption rule avoids the extra
first-wins fences required by the commutativity rule in this matrix.

\begin{table}[t]
\centering\small
\begin{tabular}{lrrrr}
\toprule
Rule & Executions & Realized & Wrong state & Successor fences\\
\midrule
Idempotence only & 15 & 14 & 5 & 0 \\
Full successor fence & 15 & 15 & 0 & 15 \\
Commutativity rule & 15 & 15 & 0 & 9 \\
Absorption rule & 15 & 15 & 0 & 6 \\
\bottomrule
\end{tabular}

\caption{Native atomic development checks. Fence counts are decisions in these
selected cells, not a production latency or scalability result.}
\label{tab:native}
\end{table}

This comparison does not imply that absorption is the weakest possible
condition. It shows why commutativity and replay neutrality are not
interchangeable obligations. No receipt-replay or model-quality benefit is
inferred from a correct point value.

\subsection{A concrete atomic-abstraction failure}
The filesystem component advertises \texttt{idempotentHint=true} for
\texttt{write\_file}. In the selected implementation, an existing file is
truncated and then written in place. Holding an old two-byte descriptor write
after its truncate, completing a retry of \texttt{A0}, and acknowledging a
newer \texttt{B-EXPANDED} write leaves \texttt{A0EXPANDED} when the old
write resumes. The result is neither the acknowledged \(A;B\) value nor an
atomic replacement with \texttt{A0}.

In the \NFilesystemCases{} primary filesystem checks, the unguarded path has
six task-state mismatches: three stale equal-length values and three mixed
unequal-length values. The physical-successor-fence path has none in its nine
executions. These observations do not make an isolated idempotence hint false;
they show that it cannot supply the stronger recovery claim being tested.
The original atomic method correctly declines a certificate for this
source profile.

The synthesized prefix test is stricter. It pauses the original attempt
before truncate, lets a retry return with \texttt{A0}, then permits only the
old truncate before observing state. In all \NPrefixCases{} executions the
settled file ultimately returns to \texttt{A0}. Nevertheless, the blind-retry
arm has \NPrefixBlindFailures{} post-acknowledgement empty observations.
The microstep rule waits for the original normal completion and coalesces
with it; it has no such acknowledged-prefix violation in its three
executions. This is the same appropriate waiting primitive offered by strict
coalescing, not a claim that the prototype beats that primitive.

\subsection{An inside-lock negative result}
All \NMemoryCases{} graph executions preserve the declared task state.
The original graph publication is delayed while already inside the real
mutation queue and file lock. The retry cannot overtake that publication, so
the intended late-old-effect schedule is not realized at this boundary.
The test does not remove the lock to manufacture a failure.

This is a scoped negative result. It does not prove that a mutex deduplicates
logical operations, that all pre-lock delays are safe, or that another
connection could never reach the mutation in a different order. It establishes
that this chosen inside-lock pause is protected by the implementation's own
serialization, without an added client fence.

\subsection{Git and document updates}
The prospectively profiled Git paths produce the results in
Table~\ref{tab:git}. With a fixed modified worktree, the original
\texttt{git\_add} can resume after an acknowledged \texttt{git\_reset} and
stage the change again. Isolated idempotence of staging has not preserved the
logical stage--reset sequence. Disjoint and same-value additions supply
successful controls; no duplicate-commit discovery is claimed.

\begin{table}[t]
\centering\small
\begin{tabular}{lrrrrr}
\toprule
Rule & Executions & Realized & Wrong state & \(B\) fences & Coalesced\\
\midrule
Idempotence only & 9 & 8 & 3 & 0 & 0 \\
Full successor fence & 9 & 9 & 0 & 9 & 0 \\
Commutativity rule & 9 & 9 & 0 & 3 & 0 \\
Absorption rule & 9 & 9 & 0 & 3 & 0 \\
In-flight coalescing & 9 & 9 & 0 & 0 & 9 \\
\bottomrule
\end{tabular}

\caption{Real Git-index checks with two tracked MCP connections. One
same-value unguarded schedule is unrealized and remains in the report.}
\label{tab:git}
\end{table}

The Chroma update paths independently reproduce a stale existing-document
write in the conflict case. Table~\ref{tab:chroma} preserves the difference
between a source-backed conditional model and a concrete-code theorem.
The profile forbids deletion and checks both IDs' presence, because the
actual update API ignores missing IDs. Neither the wrapper nor the backend is
modified to add an operation ledger.

\begin{table}[t]
\centering\small
\begin{tabular}{lrrrrr}
\toprule
Rule & Executions & Realized & Wrong state & \(B\) fences & Coalesced\\
\midrule
Idempotence only & 6 & 6 & 3 & 0 & 0 \\
Absorption rule & 6 & 6 & 0 & 3 & 0 \\
In-flight coalescing & 6 & 6 & 0 & 0 & 6 \\
\bottomrule
\end{tabular}

\caption{Real Chroma MCP updates through a separate native Chroma 1.0.16
server, with exact document reads by an independent direct SDK path.}
\label{tab:chroma}
\end{table}

\subsection{A normal browser reply is not application quiescence}
The browser binding uses an unmodified pinned Playwright MCP package and an
existing Edge executable in a new headless isolated profile. Its controlled
local page schedules a SQLite-backed HTTP mutation five seconds after a
button click. In all \NBrowserCases{} probes the normal click reply precedes
the durable application write: direct SQLite counts are zero at the reply
and one after the callback commits.

The guard receives neither the direct SQLite observation nor the callback's
private commit signal. It reports ambiguity rather than treating the normal
RPC return as completion of the application effect. This is not a
Playwright protocol violation: a successful click is not a promise that every
future application timer has finished. The result identifies a completion
binding that a stronger workflow would need to supply.

\subsection{Safety and progress with a missing original reply}
Table~\ref{tab:missing} reports the separate missing-reply cohort. The original
backend write actually completes, but its normal reply never reaches the
caller. Strict coalescing cannot supply that result and completes none of
the nine selected workflows. Idempotence-only admission completes nine but
violates three acknowledged complete goals.

The absorption rule completes the six disjoint or first-wins cases justified
by their contracts, with no acknowledged complete-goal violation. It leaves
the three conflicts ambiguous after acknowledging only \(A\).
Those are incomplete workflows, not successful recoveries. The result is an
availability distinction under these source and scope assumptions, not a
solution to arbitrary unknown commits or permanent faults.

\begin{table}[t]
\centering\small
\begin{tabular}{lrrrrr}
\toprule
Rule & Executions & Complete & Correct complete & Wrong complete & Incomplete\\
\midrule
Idempotence only & 9 & 9 & 6 & 3 & 0 \\
Absorption rule & 9 & 6 & 6 & 0 & 3 \\
In-flight coalescing & 9 & 0 & 0 & 0 & 9 \\
\bottomrule
\end{tabular}

\caption{Actual response suppression after backend completion. Injector-only
acknowledgements and delay bounds are not disclosed to the guard. Blocking is
not scored as successful recovery.}
\label{tab:missing}
\end{table}

\subsection{What the independent checker confirms}
Porcupine performs \NCheckerDecisions{} checks over the
\NCheckerHistories{} complete histories using the two explicit specifications.
The intent-aware result agrees with every task oracle. Its
\NIntentRejections{} rejections include the
\NPhysicalRejections{} histories rejected by the physical whole-tool
specification. The remaining difference is expected: a late physical
overwrite can be a legal third API call while still being an invalid repeat
of a completed logical intent.

\begin{table}[t]
\centering\small
\begin{tabular}{lrrrr}
\toprule
Source & Histories & Physical rejects & Intent-aware rejects & Oracle agreements\\
\midrule
chroma & 18 & 0 & 3 & 18 \\
filesystem & 18 & 3 & 6 & 18 \\
filesystem\_prefix & 6 & 3 & 3 & 6 \\
git & 45 & 0 & 3 & 45 \\
memory & 9 & 0 & 0 & 9 \\
native\_development & 60 & 0 & 5 & 60 \\
\bottomrule
\end{tabular}

\caption{Porcupine 1.3.1 actually executed with fair, explicit specifications.
The adapted checker agrees with the proposed task oracle; no superiority over
Porcupine is claimed.}
\label{tab:checker}
\end{table}

This check is useful precisely because it does not establish a new
linearizability algorithm. It corroborates the concrete interpretation of
the witnesses and exposes the effect of the specification chosen by the test.
The prototype adds candidate generation, source-bound materialization and
conditional recovery admission; the established checker supplies a separate
history-checking implementation.

\section{Interpretation and practical use}
An idempotence annotation should start a recovery question, not end it.
A client still needs to know which intent is repeated, which later operations
can interfere, whether the claimed effect is one step, and what evidence can
retire an outstanding physical attempt. These questions are different from
whether a JSON-RPC response was received.

The appropriate response can be an existing primitive. For the non-atomic
file profile, waiting for the old call and coalescing is justified; inventing
a backend retry cache is unnecessary. For an atomic neutral retry, waiting
for an unavailable original reply can sacrifice progress without buying
additional effect-state safety. For a conflict with unobservable completion,
the prototype cannot manufacture a fence and should say so. A real
operation-deduplication, version-check or reconciliation protocol may then be
needed. We do not claim those protocols are replaceable by this analysis.

The approach is most useful as a reviewable testing workflow around selected
write paths. A developer supplies the reference semantics and source binding;
the checker produces queries and candidate schedules; an adapter demonstrates
or rejects their realisability; the result records the exact assumptions
under which a decision follows. It is not currently a production-ready,
automatic analyser of arbitrary MCP implementations.

\section{Threats to validity and boundaries}
\paragraph{Manual models and source fidelity.}
Profiles and parameter instances were constructed by the same research process
that implemented the prototype. Independent interpreters and Porcupine
reduce implementation-error risk but do not constitute independent human
validation of those profiles. Normal return is trusted only according to the
declared source binding. Hidden backend retries, unspecified durability,
version-dependent internal behavior or a mistaken binding can invalidate
that assumption.

\paragraph{Bounds and observations.}
The microstep method covers the declared two- or three-attempt programs and
finite resource/byte footprint. It does not certify arbitrary crashes,
unbounded retries, dynamic resources, monitor restart, partial failure after
an unmodelled step, or concurrent untracked clients. Exact mutator receipts
are projected out. The atomic proofs are also conditional on their invariant
and retained context; equality of a narrower count or snapshot is not silently
substituted for that context.

\paragraph{Selection and researcher adaptation.}
The frame is purposive and small. Three components share a repository,
Chroma is an older pinned implementation, and only selected mutation paths
are exercised. Filesystem and memory became development evidence for the
microstep refinement, rather than being presented as blind validation of
that later design. Per-target models were frozen before recovery execution,
but this local procedure is not a registered experiment or an unbiased
estimate of ecosystem behavior.

\paragraph{Execution environment and failures.}
All service experiments run on one Windows machine. Repetitions check
reproducibility, not population-level failure probabilities.
\NUnrealizedPrimarySchedules{} intended primary late-effect schedules are
unrealized and retained. Redis remains unavailable. Harness failures before
exposure---including Windows path aliases, an invalid isolated-browser OS
environment and backend-startup timing---are preserved separately, not called
successful target tests. No production service or private enterprise source
is a fault-injection target.

\paragraph{What is not measured.}
We do not measure autonomous agent recovery policy quality, semantic retrieval
quality, model ranking, production incidence, solver superiority over
Servois, or deployment-scale performance. The browser application is a
controlled fixture, not a production web application. Positive model
results are not general concrete-code certificates. The study also does not
establish that a newly reported task witness is an unknown upstream defect:
many behaviors follow from established concurrency semantics and narrower
documented contracts.

\section{Related work}
RPC duplicate suppression and RIFL supply the essential prior understanding of
logical operation identity and exactly-once linearizable execution
\citep{birrell1984,rifl}. We do not rediscover their failure mode or offer a
replacement for durable server-side operation records. Earlier agent
experiments, including LIMBO, ask where repeat execution effects arise
\citep{limbo}; this work instead makes the selected recovery claim an
executable contract with explicit granularity and completion bindings.

Servois generates precise commutativity conditions, generalizes to left/right
movers and proves relative completeness \citep{servois}. Our atomic engine is
not claimed to be more expressive. Coordination avoidance and invariant
confluence likewise establish that semantic knowledge can remove unnecessary
coordination \citep{bailis}. The overlap is substantial and deliberate. The
proposed increment is the scoped tool-recovery workflow and its linkage to
real delayed effects, not a new algebra for safe concurrency.

Systematic interleaving exploration is established in CHESS
\citep{chess}; lineage-driven fault injection uses execution evidence to guide
failure exploration \citep{ldfi}. Our bounded microstep enumeration is not a
replacement for those systems or a new general concurrency-testing
algorithm. It specializes a declared retry/acknowledgement horizon and
requires source-bound adapters to validate candidate witnesses.

MCP Rig already injects dropped responses and checks real state
\citep{mcprig}. MCP-Injector studies protocol-aware faults
\citep{mcpinjector}, MCPGuard checks specification-to-SDK conformance
\citep{mcpguard}, and executable tool-contract testing addresses semantic
conformance \citep{toolcontracts}. It would be inaccurate to characterize
existing tests as uniformly schema-only or count-only. The result-only
actionability audit also makes recovery guidance an explicit research topic
\citep{actionability}. We differ in the particular contract composition:
intent-aware replay, possible interleaving inside a supposedly repeatable
tool, and source-justified completion, checked together.

Engineering systems such as Once and EffectFence already emphasize intent
identity, UNKNOWN outcomes, replay ledgers and causal fencing
\citep{once,effectfence}. Those features alone are not new here, and we did not
benchmark their implementations. Similarly, our fair Porcupine comparison
shows agreement when the same intent-aware objective is supplied, rather than
claiming that a new oracle defeats an established checker.

The contribution is therefore not a new identity law or history checker. It is
a recovery-contract workflow that carries logical identity, the selected
effect state, instruction granularity and the completion binding through
model checking, real fault materialization and action admission. The
evaluation demonstrates why discarding one of those contract parts can
change a verdict, hide an acknowledged-prefix failure, or turn correct
blocking into a falsely reported recovery.

\section{Reproducibility, disclosure and research status}
The private replication package contains the typed models, separately
implemented interpreters, source profiles, pinned dependency information,
generated queries, actual traces, normalized checker histories and table
generator. The original frame, later refinement freeze, per-target profile
freezes, unsuccessful harness attempts and unrealized schedules remain
distinguishable. Tables in this manuscript are generated from the recorded
cohorts rather than entered by hand.

The work uses public open-source components and newly created synthetic
resources. It incorporates no private enterprise source, customer data or
production workload. Browser experiments use new isolated profiles; Git
experiments disable inherited Git configuration and operate only on new
fixture repositories. Native services bind loopback and stop after use.
This manuscript has not undergone independent human peer review.

\paragraph{AI assistance and author responsibility.}
AI tools were used extensively to assist with programming, analysis and
drafting. The author takes responsibility for the content of this manuscript,
including the model/source correspondences, experimental records,
bibliography and claims. No publication-readiness certification is claimed.

\section{Conclusion}
An agent tool can be repeatable in isolation while its recovery workflow loses
an acknowledged update, exposes a partially undone file, or mistakes an RPC
reply for application completion. Testing those claims requires more than a
repeat count or an idempotence hint. The prototype checks explicitly scoped
atomic and microstep contracts, materializes selected witnesses at real
boundaries, and retains uncertainty when the required completion evidence is
missing. The resulting evidence supports a concrete recovery-testing method
within its stated bounds. It does not turn a model proof into a
production-wide exactly-once guarantee.

\bibliographystyle{plainnat}
\bibliography{references}

\begin{thebibliography}{21}
\providecommand{\natexlab}[1]{#1}
\providecommand{\url}[1]{\texttt{#1}}
\expandafter\ifx\csname urlstyle\endcsname\relax
  \providecommand{\doi}[1]{doi: #1}\else
  \providecommand{\doi}{doi: \begingroup \urlstyle{rm}\Url}\fi

\bibitem[eff()]{effectfence}
Effectfence.
\newblock Software repository.
\newblock URL \url{https://github.com/aurumflux20/effectfence}.

\bibitem[mcp()]{mcprig}
{MCP Rig}: Deterministic, {CI}-ready testing for {MCP} servers.
\newblock Software repository.
\newblock URL \url{https://github.com/gorkemgul/mcp-rig}.
\newblock Inspected commit 5c3c607c25fe94c485192c4c420894e2f64b5b87.

\bibitem[too()]{toolcontracts}
Executable tool-contract conformance testing for agentic benchmarks.
\newblock Replication repository.
\newblock URL \url{https://github.com/rohithreddybc/tool-contract-conformance}.

\bibitem[act(2026)]{actionability}
Can {MCP} clients decide what to do after failure? a result-only actionability audit.
\newblock arXiv:2609.00072, 2026.
\newblock URL \url{https://arxiv.org/abs/2609.00072}.

\bibitem[Alvaro et~al.(2015)Alvaro, Rosen, and Hellerstein]{ldfi}
Peter Alvaro, Joshua Rosen, and Joseph~M. Hellerstein.
\newblock Lineage-driven fault injection.
\newblock In \emph{Proceedings of the 2015 ACM SIGMOD International Conference on Management of Data}, pages 331--346, 2015.
\newblock \doi{10.1145/2723372.2723711}.

\bibitem[Athalye()]{porcupine}
Anish Athalye.
\newblock Porcupine: A fast linearizability checker.
\newblock Software repository.
\newblock URL \url{https://github.com/anishathalye/porcupine/tree/v1.3.1}.
\newblock Version 1.3.1; actually executed in this study.

\bibitem[Bailis et~al.(2014)Bailis, Fekete, Franklin, Ghodsi, Hellerstein, and Stoica]{bailis}
Peter Bailis, Alan Fekete, Michael~J. Franklin, Ali Ghodsi, Joseph~M. Hellerstein, and Ion Stoica.
\newblock Coordination avoidance in database systems (extended version).
\newblock arXiv:1402.2237, 2014.
\newblock URL \url{https://arxiv.org/abs/1402.2237}.
\newblock Extended version of the paper in PVLDB 8(3).

\bibitem[Bansal et~al.(2018)Bansal, Koskinen, and Tripp]{servois}
Kshitij Bansal, Eric Koskinen, and Omer Tripp.
\newblock Automatic generation of precise and useful commutativity conditions (extended version).
\newblock arXiv:1802.08748, 2018.
\newblock URL \url{https://arxiv.org/abs/1802.08748}.
\newblock Extended version of the TACAS 2018 paper.

\bibitem[Birrell and Nelson(1984)]{birrell1984}
Andrew~D. Birrell and Bruce~Jay Nelson.
\newblock Implementing remote procedure calls.
\newblock \emph{ACM Transactions on Computer Systems}, 2\penalty0 (1):\penalty0 39--59, 1984.
\newblock \doi{10.1145/2080.357392}.

\bibitem[{Chroma}()]{chroma}
{Chroma}.
\newblock Chroma {MCP} server.
\newblock Software repository.
\newblock URL \url{https://github.com/chroma-core/chroma-mcp}.
\newblock Commit 98ff67589bdcc31b730a5415ff9529433f949077.

\bibitem[Herlihy and Wing(1990)]{herlihy1990}
Maurice~P. Herlihy and Jeannette~M. Wing.
\newblock Linearizability: A correctness condition for concurrent objects.
\newblock \emph{ACM Transactions on Programming Languages and Systems}, 12\penalty0 (3):\penalty0 463--492, 1990.
\newblock \doi{10.1145/78969.78972}.

\bibitem[Iannillo et~al.(2026)Iannillo, Owotogbe, Natella, Avallone, Kudryavtseva, and Kumara]{mcpinjector}
Antonio~Ken Iannillo, Joshua~Segun Owotogbe, Roberto Natella, Francesco Avallone, Kristina Kudryavtseva, and Indika Kumara.
\newblock Tool-calling {LLM} agents under realistic {MCP} faults: Scale does not purchase fault tolerance.
\newblock arXiv:2610.32555, 2026.
\newblock URL \url{https://arxiv.org/abs/2610.32555}.

\bibitem[Lee et~al.(2015)Lee, Park, Kejriwal, Matsushita, and Ousterhout]{rifl}
Collin Lee, Seo~Jin Park, Ankita Kejriwal, Satoshi Matsushita, and John Ousterhout.
\newblock Implementing linearizability at large scale and low latency.
\newblock In \emph{Proceedings of the 25th Symposium on Operating Systems Principles}, 2015.
\newblock \doi{10.1145/2815400.2815416}.
\newblock URL \url{https://web.stanford.edu/~ouster/cgi-bin/papers/rifl.pdf}.

\bibitem[Li(2026)]{limbo}
Jiapeng Li.
\newblock Where does exactly-once live? model, harness, and tool-contract effects on duplicate side effects in {LLM} agents.
\newblock arXiv:2609.29095, 2026.
\newblock URL \url{https://arxiv.org/abs/2609.29095}.

\bibitem[{Microsoft}()]{playwright}
{Microsoft}.
\newblock Playwright {MCP}.
\newblock Software repository.
\newblock URL \url{https://github.com/microsoft/playwright-mcp}.
\newblock Commit 3afdae855f62998c87644144d4f7dc705ac24dcf; playwright-core 1.66.0-alpha-2026-11-09.

\bibitem[{Model Context Protocol}({\natexlab{a}})]{mcpschema}
{Model Context Protocol}.
\newblock Schema reference, {\natexlab{a}}.
\newblock URL \url{https://modelcontextprotocol.io/specification/2025-11-25/schema}.
\newblock Specification dated 25 November 2025; accessed 10 November 2026.

\bibitem[{Model Context Protocol}({\natexlab{b}})]{reference}
{Model Context Protocol}.
\newblock Reference servers.
\newblock Software repository, {\natexlab{b}}.
\newblock URL \url{https://github.com/modelcontextprotocol/servers}.
\newblock Commit 36380287ed3b15d115b55f546617d723072b1296.

\bibitem[Musuvathi et~al.(2007)Musuvathi, Qadeer, and Ball]{chess}
Madan Musuvathi, Shaz Qadeer, and Thomas Ball.
\newblock {CHESS}: A systematic testing tool for concurrent software.
\newblock Technical Report MSR-TR-2007-149, Microsoft Research, 2007.
\newblock URL \url{https://www.microsoft.com/en-us/research/publication/chess-a-systematic-testing-tool-for-concurrent-software/}.

\bibitem[{Once}()]{once}
{Once}.
\newblock {MCP} retry safety: Safe retries for state-changing tools.
\newblock Engineering documentation.
\newblock URL \url{https://onceexec.com/mcp-retry-safety/}.

\bibitem[{Redis}()]{redis}
{Redis}.
\newblock Redis {MCP} server.
\newblock Software repository.
\newblock URL \url{https://github.com/redis/mcp-redis}.
\newblock Commit f7bd20a2bfd4f4183519db61f4bcd2a905761549.

\bibitem[Sun et~al.(2026)Sun, Liu, Dang, Zheng, Yang, and Liu]{mcpguard}
Xikai Sun, Kebin Liu, Fan Dang, Yawen Zheng, Zihao Yang, and Yunhao Liu.
\newblock {MCPGuard}: Finding conformance gaps in {MCP SDKs} via dual-track analysis and dynamic adjudication.
\newblock arXiv:2611.03662, 2026.
\newblock URL \url{https://arxiv.org/abs/2611.03662}.

\end{thebibliography}

\appendix
\section{A replayable acknowledged-prefix witness}
\begin{table}[ht]
\centering\small
\begin{tabular}{p{0.09\linewidth}p{0.48\linewidth}p{0.29\linewidth}}
\toprule
Order & Observed event & File state\\
\midrule
1 & Old \(A\) opens the existing file; pause before truncate & \texttt{BASE}\\
2 & Client deadline; old physical call is not cancelled & \texttt{BASE}\\
3 & Retry truncates, writes, closes and returns normally & \texttt{A0}\\
4 & Old truncate executes; pause before its write & empty\\
5 & Independent read while the old reply is still pending & empty\\
6 & Old write resumes and returns normally & \texttt{A0}\\
\bottomrule
\end{tabular}
\caption{The synthesized two-attempt witness. A settled-state-only oracle
passes at order 6, but the acknowledged-prefix objective fails at order 5.}
\end{table}
The solver witness and concrete trace have different roles. The former shows
the behavior allowed by the declared program; the latter establishes the
particular source binding and independent observation. The runtime decision
is made from the program and allowed completion capability, not from the
injector's knowledge that the old call is paused at order 1 or 4.

\section{Conditions attached to every positive result}
A model result is used only with its declared resource and intent mapping,
state signature, feasible invariant, preserved context and effect boundaries.
An atomic result assumes an atomic effect. A bounded microstep result assumes
the declared instruction granularity and attempt horizon. Strong completion
must come from the specified binding; client timeout and unsuccessful
transport are not substitutes. Exact receipt replay, external writers,
untracked backend retries and monitor persistence require additional
contracts. If those conditions cannot be established, the appropriate output
is uncertainty or a narrower test claim, not a silent PASS.
\end{document}